\documentclass[10pt,conference]{IEEEtran}
\IEEEoverridecommandlockouts
\usepackage{cite}
\usepackage{amsmath,amssymb,amsfonts}
\usepackage{algorithm}
\usepackage[noend]{algpseudocode}
\usepackage{caption}
\usepackage{subcaption}
\usepackage{graphicx}
\usepackage{textcomp}
\usepackage{dcolumn}
\usepackage{bm}
\usepackage{braket}
\usepackage{hyperref}
\def\BibTeX{{\rm B\kern-.05em{\sc i\kern-.025em b}\kern-.08em
    T\kern-.1667em\lower.7ex\hbox{E}\kern-.125emX}}
    
\usepackage[svgnames, table]{xcolor}
\usepackage{tabularx, makecell, linegoal}

\DeclareMathOperator*{\argmax}{arg\,max}

\usepackage{ifthen}

\newboolean{ShowComments}
\setboolean{ShowComments}{true}  
\ifthenelse{\boolean{ShowComments}}%
	{
		\newcommand{\ColorComment}[3]{%
				{\colorbox{#1}{\color{White}   \textsf{\textbf{#2}}} \textcolor{#1}{#3}}}

	}%
	{
		\newcommand{\ColorComment}[3]{}

	}%

\definecolor{rdvcolor}{rgb}{0,0.5,0}
\definecolor{satohcolor}{RGB}{254,0,0}
\definecolor{michalcolor}{RGB}{255,127,80}
\definecolor{naphanncolor}{RGB}{112, 51, 173}

\begin{document}

\title{Amplitude Amplification for Optimization\\ via Subdivided Phase Oracle}

\author{
\IEEEauthorblockN{Naphan Benchasattabuse\IEEEauthorrefmark{1}\IEEEauthorrefmark{4}, 
Takahiko Satoh\IEEEauthorrefmark{2}\IEEEauthorrefmark{4}, 
Michal Hajdu\v{s}ek\IEEEauthorrefmark{1}\IEEEauthorrefmark{4}, 
and Rodney Van Meter\IEEEauthorrefmark{3}\IEEEauthorrefmark{4}}\\ 

\IEEEauthorblockA{\IEEEauthorrefmark{1}\textit{Graduate School of Media and Governance, Keio University Shonan Fujisawa Campus, Kanagawa, Japan}}
\IEEEauthorblockA{\IEEEauthorrefmark{2}\textit{Graduate School of Science and Technology, Kanagawa, Japan}}
\IEEEauthorblockA{\IEEEauthorrefmark{3}\textit{Faculty of Environment and Information Studies, Keio University Shonan Fujisawa Campus, Kanagawa, Japan}}
\IEEEauthorblockA{\IEEEauthorrefmark{4}\textit{Quantum Computing Center, Keio University, Kanagawa, Japan}\\
\{whit3z,satoh,michal,rdv\}@sfc.wide.ad.jp}}

\maketitle
\thispagestyle{plain}
\pagestyle{plain}

\begin{abstract}
We propose an algorithm using a modified variant of amplitude amplification to solve combinatorial optimization problems via the use of a subdivided phase oracle. 
Instead of dividing input states into two groups and shifting the phase equally for all states within the same group, the subdivided phase oracle changes the phase of each input state uniquely in proportion to their objective value. 
We provide visualization of how amplitudes change after each iteration of applying the subdivided phase oracle followed by conventional Grover diffusion in the complex plane.
We then show via numerical simulation that for normal, skew normal, and exponential distribution of objective values, the algorithm can be used to amplify the probability of measuring the optimal solution to a significant degree independent of the search space size.
In the case of skew normal and exponential distributions, this probability can be amplified to be close to unity, making our algorithm near deterministic.
We then modify our algorithm in order to demonstrate how it can be extended to a broader set of objective value distributions.
Finally, we discuss the speedup compared to classical schemes using the query complexity model, and show that our algorithm offers a significant advantage over these classical approaches.

\end{abstract}

\begin{IEEEkeywords}
Quantum computing, Grover's search algorithm, optimization, amplitude amplification
\end{IEEEkeywords}

\section{Introduction}
\label{sec:introduction}

Grover's search algorithm~\cite{grover-alg} is one of the most celebrated quantum algorithms which demonstrated the computational power of quantum computers that classical computers cannot reach.
The problem that Grover's algorithm solves can be explained as finding one of the $t$ inputs among $N$ total inputs that matches some certain constraints, $f(x) = 1$, given by an oracle function $f(x)$.
The assumption about the oracle is that nothing is known about its construction or whether the problem has any structure we can exploit to infer the solution from previously seen inputs and outputs.
Grover's algorithm can accomplish this task in just $O(\sqrt{N/t})$ queries, compared to the classical approach which requires $O(N/t)$ queries as we need to exhaustively check all inputs.

Since the original algorithm proposed by Grover in 1996, there have been multiple advancements to the algorithm.
It is shown that Grover's algorithm is a special case, and is optimal~\cite{boyer-tight-bounds, bennett-strength-weakness, morales-variationally-learning-grover}, of a more general approach, later known as ``amplitude amplification'' technique~\cite{brassard-bhmt-amplitude-amplification-estimation, farhi-analog-analogue, farhi-adiabatic-grover-optimal, childs-quantum-walk-search, biham-arbitrary-initial-grover}.
The use of a non-$\pi$ phase shift in place of the usual $\pi$-flip of Grover oracle was investigated in~\cite{long-arbitrary-phase-grover, long-phase-matching-search, hoyer-arbitrary-phase-aa, biham-general-grover} and later generalized into a technique to increase the success probability of Grover to unity~\cite{long-grover-zero-failure, toyama-grover-certainty}. 
It also finds applications in solving the souffle problem, too few or too many iterations deteriorates the success probability, which is known as the fixed-point searching~\cite{yoder-fix-point-search,  grover-fix-point-search, grover-fix-point-search-examples, kwon-qaao-fix-point-grover}.

In recent years, there have been investigations on the viability of using amplitude amplification technique with a new variant of non-trivial phase shift oracle, one that does not simply recognize two classes of inputs but rather acts on each input uniquely depending on some cost functions~\cite{satoh-spo, shyamsundar-nbaa, koch-gaussian-amplitude-amplification}.
In this paper, we aim to extend the use of this non-trivial phase shift oracle to solve combinatorial optimization problems, instead of the usual adaptive Grover's algorithm~\cite{durr-hoyer-min-finding, gilliam-adaptive-grover, gilliam-binomial-grover} where we try to find better solutions by running the algorithm repeatedly until we cannot find a better solution.

The contribution of this paper is to introduce the use of a non-trivial phase shift oracle which we refer to as a \emph{subdivided phase oracle} (SPO)~\cite{satoh-spo}, replacing the canonical Grover oracle in the amplitude amplification process, to solve combinatorial optimization problems.
We show that this approach can be used to efficiently find the solutions of a large class of optimization problems, characterized by a number of objective functions.

We describe the setting of maximization problems (section~\ref{sec:optimization}) which we then discuss one approach to construct a subdivided phase oracle operator from the given objective function and how the objective values can be embedded into the phase of each basis state (section~\ref{sec:oracle}).
We define our algorithm and provide visualization of how amplitudes change after each iteration of applying the subdivided phase oracle followed by conventional Grover diffusion (inverse around the average).
This modified amplitude amplification process is complex and non-trivial to mathematically derive closed form expressions. 
Instead, we give intuitions on how it can amplify the probability of measuring the solution states and discuss the relationship between the distribution of objective values, oracle construction, and performance of the algorithm (section~\ref{sec:spo-amplitude-amplification}). 
We then analyze the performance of the algorithm and show results via simulations of applying it to several classes of objective functions, plotting the probability of measuring the solutions with respect to the number of iterations (section~\ref{sec:results}).
We also discuss a dynamic approach where the oracle operator changes at each iteration depending on current amplitude values, although this approach is purely theoretical since it is impossible to keep track of all amplitudes in large space, it suggests that a heuristic or variational~\cite{cerezo-vqa} approach may exist in selecting parameters to construct the oracle. 
We then propose a sampling scheme based on the modified amplitude amplification process, plot the trade-off between the number of sampling trials and iterations, and derive conditions to which the sampling scheme gives a better expected run time than classical sampling or exhaustive search (section~\ref{sec:advantage}).
We end the paper by discussing possible ways of extending the ideas we present here, pointers to efficiently realize the subdivided phase oracle in circuit models, connections to QAOA~\cite{hadfield-qaoa-redefined, bartschi-grover-mixer-qaoa, farhi-qaoa} and pointers to make it variational, and utilizing the visualization we developed to create a more suitable diffusion operator to increase the performance of our algorithm (section~\ref{sec:conclusion}).

\section{Optimization Problem Setting}
\label{sec:optimization}

The problem we consider is the combinatorial optimization problem.
For example, ``what is the shortest route starting from city $A$ to city $B$ given a map of roads currently available?''. 
An optimization problem can be described by three elements; a set of all possible inputs $X$, an objective function $f(x)$ defined over part of the input space, and a Boolean function $g(x)$ which checks for the feasibility of the input.
Using the above example, $X$ refers to all possible routes connecting $A$ and $B$, $f(x)$ refers to the distance of each route, and since some roads might be closed due to constructions or accidents, $g(x)$ refers to the state of whether a route can be used at the time of travel.
The task is to find an input $x \in X$ such that $f(x)$ is the maximum or the minimum while $g(x) = 1$,
\begin{equation}
\label{eq:optimization}
    \argmax_{x \in X} f(x) \text{ such that } g(x) = 1.
\end{equation}

In this study, we will focus on the combinatorial optimization problem, where the input is defined over the set of binary string $x = \{0,1\}^n$ of length $n$, which gives $N = 2^n$ possible inputs.
The properties of problems we consider in this work and their objective functions are as follow:
\begin{itemize}
    \item We will only study the task of maximizing $f(x)$. 
    Since the minimization can be transformed into maximization by multiplying every objective value by $-1$.
    
    \item We assume that all $N$ inputs are feasible. 
    Since the constraint function $g(x)$ can be absorbed into $f(x)$ by transforming $f(x)$ into $f'(x) = f(x) - (1 - g(x))C$, where $C$ is a positive constant chosen such that any feasible input has higher objective value than all unfeasible inputs ($f'(x) > f'(y)$ for all $x, y$ when $g(x) = 1$ and $g(y) = 0$).
    
    \item We assume the range of $f(x)$ are non-negative real values.
    Since we are considering finite search space, the objective values are bounded and minimum value exists. 
    Then we transform $f(x)$ to $f'(x)$ such that $f'(x) = f(x) + \left|f(x_{min})\right|$ where $f(x_{min})$ is the minimum value.
\end{itemize}
%





\section{Subdivided Phase Oracle}
\label{sec:oracle}


Here in place of the two-class recognition oracle $O_G$ which is used in amplitude amplification for canonical Grover, we will define $O_\phi$ which changes the phase of each input in proportion to its objective value given by the objective function $f(x)$.
Even though the word ``oracle'' normally refers to a black-box Boolean function which recognizes solutions to decision or search problems, we will keep using it to refer to this phase shift operator to reflect the modification from amplitude amplification technique, and to be consistent with~\cite{satoh-spo, shyamsundar-nbaa, koch-gaussian-amplitude-amplification}.
We will refer to the oracle used in this paper as the \emph{subdivided phase oracle} or just the phase oracle, as taken from~\cite{satoh-spo}, since we \emph{subdivided} the objective value into the phase of each state with equal proportion.
In this section we will explain how one could construct a subdivided phase oracle from a given objective function.


The action of the phase oracle $O_{\phi}$ is to embed the objective value into the phase of each computational basis state,
%
\begin{equation}
\label{eq:phase_oracle}
    O_\phi \ket{x} \rightarrow e^{i \phi(x)}\ket{x},
\end{equation}
where $\phi(x)$ is a function that maps the set of possible inputs $\{ 0, 1, 2, \cdots, N-1 \}$ to the phase angle with which each state is shifted in the range $[ \theta, 2\pi + \theta ]$.
The mapping from objective function $f(x)$ to $\phi(x)$ is done by
\begin{equation}
    \label{eq:phi_mapping}
    \phi(x) = k f(x) + \theta,
\end{equation}
where $k$ is a parameter we can adjust and $\theta$ is the minimum objective value.
The action of $\theta$ is to introduce a global phase which can be ignored, meaning the action of the phase oracle can be written as follows,
%
\begin{equation}
\label{eq:spo_combined}
    O_\phi(k) \ket{x} \rightarrow e^{i k f(x) }\ket{x}.
\end{equation}
We make a distinction between $f(x)$ and $\phi(x)$ so that $f(x)$ is always referred to the objective function while $\phi(x)$ is the mapped function that will be used to perform the phase shift to the computational bases.
When the context is clear which $k$ is used or when it is not important to denote the selected $k$ value, we will omit the $k$ and write the oracle as just $O_\phi$.


From Eq.~(\ref{eq:spo_combined}), it can be seen that the action of the oracle is invariant under shift of the objective values.
Furthermore, the nature of the oracle does not depend nor exploit the structure of the problem, which leads to the following two simplifications:

\begin{itemize}
    \item $f(x)$ can be taken to be a non-decreasing function, $f(x) \leq f(y)$ for all $x < y$, since the action of oracle is invariant under input permutation.
    
    \item We take the minimum objective value to be $0$, $f(x_{min}) = 0$. As we have shown in Eq.~(\ref{eq:phi_mapping}) and Eq.~(\ref{eq:spo_combined}) that the subdivided phase oracle is invariant under shift to the objective values.
\end{itemize}

\section{Amplitude Amplification using phase oracle}
\label{sec:spo-amplitude-amplification}

Having established the problem definition, we now present our algorithm as below.

\begin{algorithm}
\caption{Optimization Amplitude Amplification}
\label{alg:algorithm-1}
\begin{algorithmic}[1]
\State Initialize state $\ket{s} = \ket{0}^{\otimes n}$
\State Prepare uniform superposition $\ket{\Psi_0} \gets H^{\otimes n}\ket{s}$
\For{$j \approx \sqrt{N}$ times}
    \State Apply $O_\phi (k)$ (Subdivided Phase Oracle)
    \State Apply $D$ (Diffusion Operator)
    \State $\ket{\Psi_{j}} \gets D O_\phi (k) \ket{\Psi_{j-1}}$ 
\EndFor
\State Measure the qubit state $\ket{\Psi}$
\end{algorithmic}
\end{algorithm}
It has the same structure as Grover's algorithm but uses the subdivided phase oracle $O_\phi(k)$ in place of the binary oracle operator.
The diffusion operator is given by
\begin{equation}
    D = 2|\Psi_0\rangle\langle\Psi_0| - I,
\end{equation}
where $I$ is the identity operator.

\subsection{Visualizing the effect of subdivided phase oracle}
\label{subsec:visualization}
\begin{figure*}[htb]
    \centering
    \includegraphics[width=0.7\textwidth]{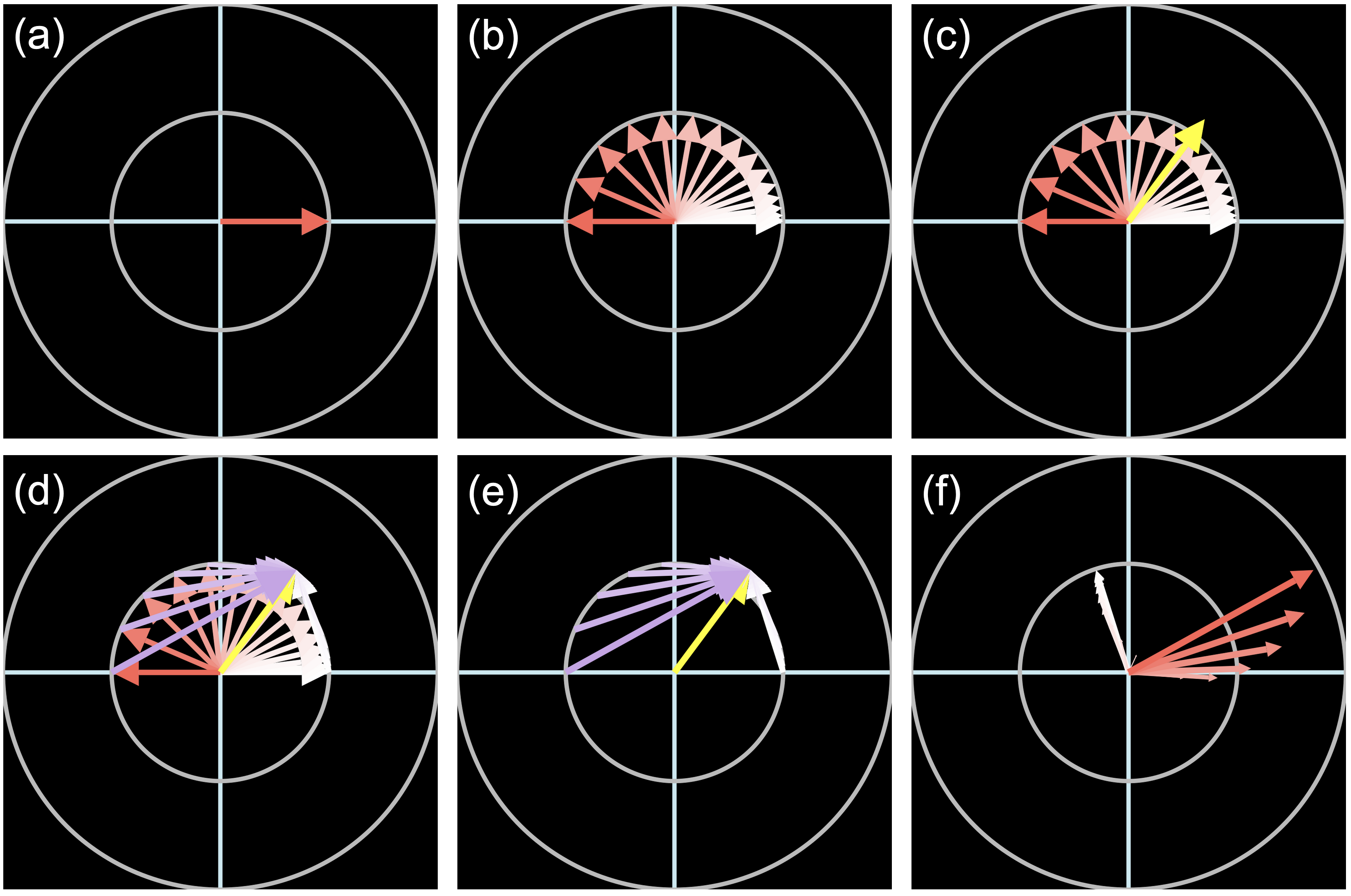}
    \caption{2D complex plot of how amplitudes of states (N = 16) change during amplitude amplification using $f(x) = x^2$ and $k = \pi / (N-1)^2$; The colors of ranging from white to red (and white to purple in (c)-(d)) denotes the objective value. (a) Amplitudes at the time of initialization after making uniform superposition. (b) Amplitudes after applying the phase oracle $O_\phi$. (c) Same time step as (b) but depicting a yellow vector which denotes the mean amplitude multiplied by 2. (d) and (e) purple vectors denote the difference from the amplitude of each state to the doubled mean vector which is equal in size and directions to amplitudes after diffusion operator applied. (f) Amplitudes after one iteration of phase oracle and diffusion operator applied.}
    \label{fig:spo-evolution}
\end{figure*}
Before applying our algorithm to various optimization problems, it is in order to gain visual intuition about how it leads to amplitude amplification.
Unlike the Grover's algorithm, where all vector amplitudes are real numbers, the subdivided phase oracle produces complex amplitudes.
Therefore each amplitude corresponding to a computational basis state $|x\rangle$ can be represented as a vector in the complex plane. 

Figure~\ref{fig:spo-evolution}(a) represents the amplitudes of the equal superposition state $|\Psi_0\rangle$, where all vectors are real and have the same length given by $1/\sqrt{N}$, where $N=2^n$.
The action of the phase oracle $O_{\phi}(k)$ is to rotate each vector corresponding to all the computational basis states around the origin as seen in Figure~\ref{fig:spo-evolution}(b).
This introduction of relative phases does not change the probabilities of measuring the individual computational basis state.
The amplitude amplification occurs after the application of the diffusion operator $D$, which can be visualized in the following three steps:
%
\begin{enumerate}
    \item Find the mean amplitude vector and scale its length by two, represented by a yellow arrow in Figure~\ref{fig:spo-evolution}(c).
    \item Compute the vector differences between the amplitude vectors and the scaled mean amplitude vector, depicted by purple arrows in Figures~\ref{fig:spo-evolution}(d) and (e).
    \item Finally, replace the old amplitude vectors with the newly computed vectors from Step 2. This effectively shifts these updated amplitude vectors to the origin, as shown in Figure~\ref{fig:spo-evolution}(f).
\end{enumerate}
The vector corresponding to the probability amplitude of the optimal computational basis state representing the solution to the optimization problem is given by the most saturated red color in Fig.~\ref{fig:spo-evolution}.
We can see that application of the subdivided phase oracle followed by the diffusion operator increases its magnitude, amplifying the probability of measuring the optimal solution.

\subsection{Amplifiability via subdivided phase oracle}
\label{subsec:amplifiability}

Grover oracle, when viewed in the same picture can be tracked easily, as the $\pi$ phase flip would make the group of marked state's amplitude vectors point to the reverse direction.
When the diffusion operator is applied, the mean would lie on the real number line and after the difference vectors are created, all amplitude vectors would still lie on the real number line.
It is noteworthy that when everything stays on the real number line, the amplitude of the marked states only gets larger when the mean vector is pointing in the opposite direction.
The mean vector would get smaller and smaller after each iteration until it stops pointing in the opposite direction and starts to point in the same direction as vectors of marked states.

Sequences of applying $O_\phi(k)$ followed by the diffusion operator lead to a more complex scenario than the Grover oracle, as both the directions and norm of the amplitude vectors and the mean vectors largely depend on the choice of $O_\phi(k)$.
The visualization that we discussed in the previous subsection however offers an intuitive picture when and how the subdivided phase oracles leads to amplitude amplification.
Since we are interested in using this algorithm to solve optimization problems, we are only interested in the optimal state vector.
The optimal state solution will grow larger up to a certain point and then shrink back and repeat this process in cycles.
The condition when the vector gets larger after diffusion is only 
when $ \left|\left| 2\overrightarrow{\alpha}_{mean} - \overrightarrow{\alpha}_{best} \right|\right| > \left|\left| \overrightarrow{\alpha}_{best} \right|\right|$ where $\overrightarrow{\alpha}$ denotes the amplitude vector.
This means the angle between the mean vector and the amplitude vector of the optimal state $\theta(\overrightarrow{\alpha}_{best}, \overrightarrow{\alpha}_{mean})$ should be between $\pi/2$ and $3\pi/2$ in order to observe significant amplitude amplification after every iteration of our algorithm.
For angles outside this range, amplitude amplification occurs but its magnitude is only marginal.
As we will see later, the choice of $k$ affects this angle and hence the level of amplitude amplification.


\section{Results}
\label{sec:results}


Before we dive into the simulations and their results, we first need to discuss the connections between the objective function $f(x)$ and the distribution of its mapped objective values.
Recall that we want to have the angle between the best state amplitude vector and the mean vector be as close as opposite direction as possible, this success condition of the amplification relies heavily on the distribution of the objective values.
Hence we will categorize the sets of simulations based on the \emph{density} of the objective function.

In addition to the normal, skew normal, and exponential distributions, we consider the case where the objective function follows basic classes of polynomials and exponential.
Although these objective functions are artificially generated and probably will not come up from natural real-world problems, we include them in our study to look at a more general behavior of the algorithm.

\subsection{Normally distributed objective values}
\label{subsec:normal-like-dist}

\begin{figure}[htb]
    \centering
    \includegraphics[width=\columnwidth]{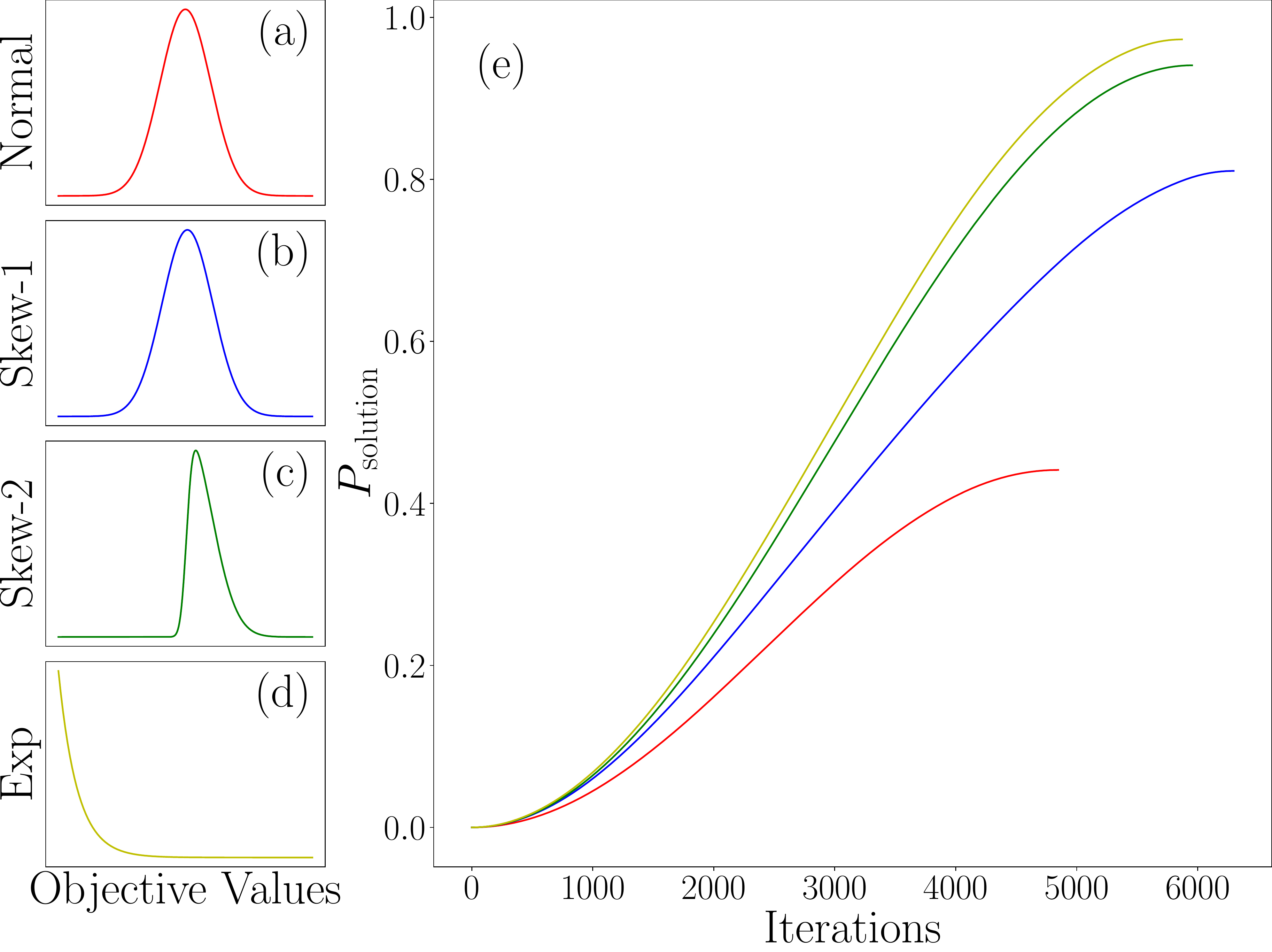}
    \caption{Amplitude amplification for various distributions of objective values with search space $N \approx 2^{25}$. (a) Normal distribution with $\sigma = 10$, (b) skew normal distribution with $\sigma = 10$, $\alpha=0.1$ and (c) $\alpha=5$, and (d) exponential distribution with all become amplified to varying degrees as seen in (e).}
    \label{fig:prob-various-dist}
\end{figure}

We begin by considering the objective values to be normally distributed according to the probability density function,
\begin{equation}
    \phi_{\text{norm}} (x) = \frac{1}{\sigma \sqrt{2\pi}} e^{-\frac{(x-\mu)^2}{2\sigma^2}},
    \label{eq:normal_dist}
\end{equation}
where $\mu$ denotes the mean and $\sigma$ the standard deviation of the distribution, as shown in Fig.~\ref{fig:prob-various-dist}(a).
This is a natural assumption as many instances of NP-hard optimization problems can be transformed into the Ising problem~\cite{lucas-ising-formulation-np}.
Such problems can be described by using local interactions between a small number of bits of the entire bit string.
This property in turn allows for the distribution of objective values of a randomly generated large instance of such NP-hard problems with local cost term sampled from a single distribution to be well approximated by a normal distribution.
This approximation becomes more accurate with increasing size of the problem due to the central limit theorem and the law of large numbers.

Normal distribution is symmetric, which makes it possible for the phase difference between the mean amplitude vector and the optimal solution to be $\pi$, leading to excellent amplification results.
In fact, using the visualization technique discussed in subsection~\ref{subsec:visualization}, it is straightforward to observe that the mean amplitude vector and the vector corresponding to the optimal solution $|N-1\rangle$ can be made to always lie on the real axis for appropriately chosen $k$.

%
Choosing $k$ such that the mean amplitude vector and the solution vector to lie on the real axis is actually the optimal $k$ value. 
Since the normal distribution is symmetric, it is the case that applying the algorithm does not only amplify the best state $\ket{N-1}$ but it also amplify the worst state $\ket{0}$ to the same probability.


\subsection{Skew normal distribution}
\label{subsec:skewed-normal-dist}

We have demonstrated that optimization problems with normally distributed objective values can be tackled easily by our algorithm.
The key property that made this possible was the symmetry of the normal distribution.
Real-world optimization problems on the other hand are unlikely to possess such nice property.
In this and the following subsections, we evaluate the performance of our algorithm for optimization problems with objective value distributions which are not symmetric.
We show that not only do they not pose an obstacle to our algorithm, asymmetry in the distributions works in our favor.

To capture the deviation from normally distributed objective values we now turn to \emph{skew normal distributions},
\begin{equation}
    \phi_{\text{skew}}(x;\alpha) = 2 \phi_{\text{norm}}(x)\Phi(x;\alpha),
    \label{eq:skew_dist}
\end{equation}
where $\phi_{\text{norm}}(x)$ is the probability density function of the normal distribution given in Eq.~(\ref{eq:normal_dist}), and
\begin{equation}
    \Phi(x;\alpha) = \frac{1}{\sqrt{2\pi}} \int_{-\infty}^{\alpha(x-\mu)/\sigma} e^{-t^2/2}dt.
\end{equation}
The parameter $\alpha$ controls the shape of the distribution, as seen in Fig.~\ref{fig:prob-various-dist}(b)-(c).
For $\alpha=0$, $\phi_{\text{skew}}(x;\alpha)$ becomes the normal distribution of Eq.~(\ref{eq:normal_dist}), while for $|\alpha|>0$, the distribution in Eq.~(\ref{eq:skew_dist}) becomes asymmetric.
In the limit of $\alpha\rightarrow\infty$, $\phi_{\text{skew}}(x;\alpha)$ becomes a half-normal distribution.

We observe that as for the case of a normal distribution, there is always an optimal parameter $k$ which amplifies the amplitude of the optimal solution to a significant degree as depicted in Fig~\ref{fig:prob-various-dist}(e).

\subsection{Exponential distribution}
\label{subsec:expo-dist}

We have seen that our algorithm can be applied to optimization problems involving normal and skew normal distributions of objective values.
In order to assess the applicability of the algorithm further, we expand our discussion to \emph{exponential distribution},
\begin{equation}
    \phi_{\text{exp}} (x; \lambda) = \lambda e^{-\lambda x},
\end{equation}
where $\lambda$ is the rate parameter, as depicted in Fig.~\ref{fig:prob-various-dist}(d).
Despite their rather different definitions, all three objective distribution functions described above share a number of important features when it comes to amplitude amplification, which we discuss in the following subsection.

%
%

\subsection{Common observations}
\label{subsec:common_obs}

\begin{figure}[htb]
    \centering
    \includegraphics[width=\columnwidth]{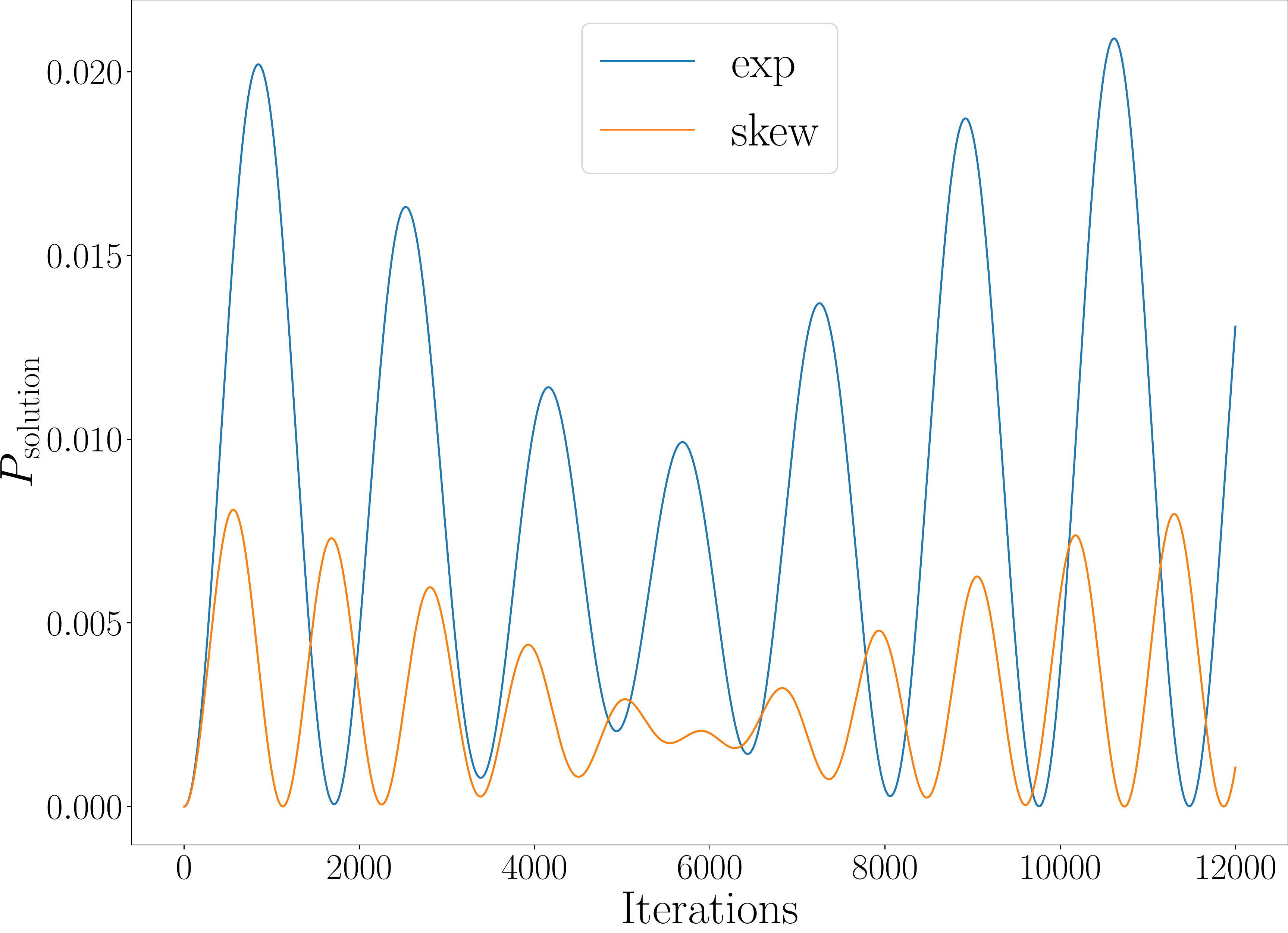}
    \caption{Evolution of the probability of the optimal solution $P_{\text{solution}}$ as a function of the iterations of Algorithm~\ref{alg:algorithm-1}. The shape of the plots are ubiquitous across all objective functions and choice of $O_\phi(k)$.}
    \label{fig:general-trend-sine}
\end{figure}
We have seen how Algorithm~\ref{alg:algorithm-1} can be applied to optimization problems with normal, skew normal, and exponential distribution of objective values.
In Fig.~2(e), we display how the probability of measuring the optimal value, $P_{\text{solution}}$, gets amplified with every successive application of the subdivided phase oracle $O_{\phi}(k)$ and the diffusion operator $D$.
Initially, $P_{\text{solution}}=1/N\approx0$, as the input size of the optimization problem is set to $N=2^{25}$.
We observe that the type of distribution affects both how many iterations are needed to achieve the maximum amplitude amplification, given by $P_{\text{solution}}^*$, as well as the value of $P_{\text{solution}}^*$.
Interestingly, when the objective values are distributed according to the normal distribution, Algorithm~\ref{alg:algorithm-1} yields $P_{\text{solution}}^*\approx0.5$.
For the case of the two asymmetric objective value distributions on the other hand, the probability of measuring the optimal value gets amplified more strongly.
$P_{\text{solution}}^*=0.94$ for the skew normal distribution, and $P_{\text{solution}}^*=0.97$ for the exponential distribution.

Figure~2(e) displays the number of iterations needed to achieve maximum amplification of the optimal solution.
Extending the number of iterations further reveals that $P_{\text{solution}}$ begins to oscillate as depicted in Fig.~\ref{fig:general-trend-sine}.
This is reminiscent of the behavior observed in amplitude amplification using the conventional binary oracle.
Unlike the binary oracle, the oscillations of $P_{\text{solution}}$ display amplitude modulation not observed previously, leading to a number of oscillations of $P_{\text{solution}}$ before the value of $P_{\text{solution}}^*$ is reached again.
This beating behavior is a consequence of the fact that the subdivided phase oracle introduces complex probability amplitudes into the state vector $|\Psi_j\rangle$.
This basic pattern is observed for all distributions and tested objective functions.
Increasing $k$ in the subdivided phase oracle $O_{\phi}(k)$ leads to decreased frequency of these oscillations.

\begin{figure*}
    \includegraphics[width=\textwidth]{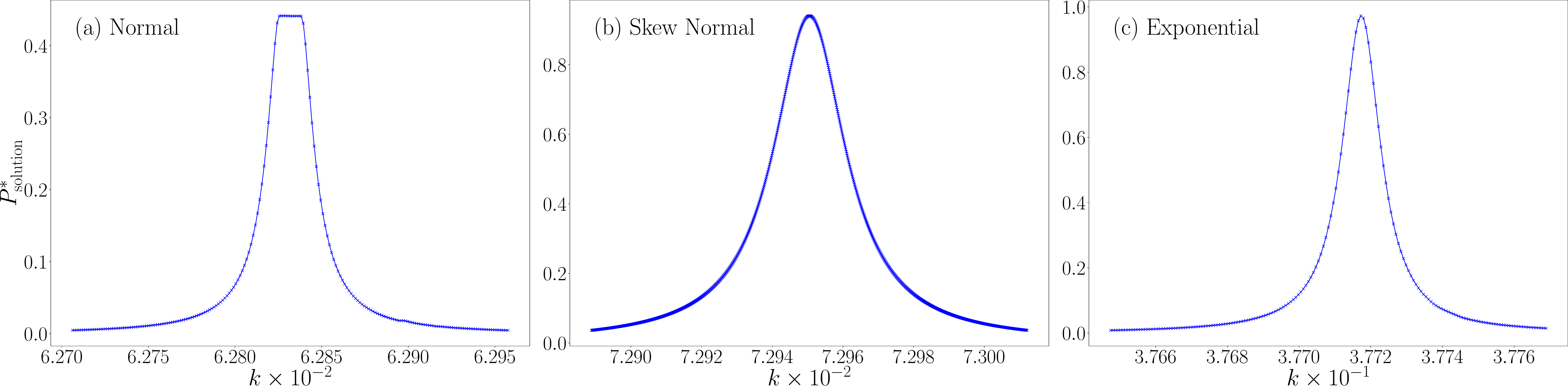}
    \caption{Sensitivity of the performance of Algorithm~\ref{alg:algorithm-1} in terms of the scaling of $P_{\text{solution}}^*$ with varying $k$. (a) Normal distribution, (b) skew normal, and (c) exponential distribution all display extreme sensitivity when deviating from the optimal $k$.}
    \label{fig:angle-sensitivity}
\end{figure*}
As discussed in Section~\ref{sec:spo-amplitude-amplification}, in order to obtain the maximum probability $P_{\text{solution}}^*$, we have to choose the optimal $k$ in the subdivided phase oracle $O_{\phi}(k)$.
Algorithm~\ref{alg:algorithm-1} is sensitive to this choice and deviating from the optimal $k$ rapidly decreases the achievable amplitude amplification.
In Fig.~\ref{fig:angle-sensitivity}, we show how the maximum amplified probability $P_{\text{solution}}^*$ depends on the deviation of $k$ from its optimal value for normal and skew normal distributions.

In Grover's algorithm or the usual amplitude amplification algorithm, if the diffusion operator is kept the same, being the inversion around the mean operator $2\ket{\Psi_0}\bra{\Psi_0} - I$, it is known that in large search space, only $\pi$ phase shift on the solution states are acceptable.
And if one were to change the oracle to use phase angle other than $\pi$, one would need to change the diffusion operator to match the angle of the oracle under some constraints as explored in~\cite{long-arbitrary-phase-grover, hoyer-arbitrary-phase-aa}.
The same effect can also be seen in our results in Fig~\ref{fig:general-trend-sine}.


The sensitivity to the angle or the $k$ value is not just for the efficiency in terms of number of iterations but is crucial in order to make algorithm~\ref{alg:algorithm-1} work.

An important observation to make at this point is that the effectiveness of our algorithm is not affected by the size of the input space.
This is a desired feature and signifies the applicability of our subdivided phase oracle approach to optimization even for extremely large data sets.
Below, we discuss optimization problems which do not share this property and the degree of amplitude amplification diminishes with increasing input space.
This will lead to a modified algorithm which overcomes this issue.

\subsection{Overcoming size-induced limits to amplifiability}
\label{subsec:size-induced-limits}

We have seen above that Algorithm~\ref{alg:algorithm-1} possesses a number of desired properties when applied to common distributions of objective values.
One such property is that its effectivity does not diminish with increasing size of the input space of the optimization problem.
In this subsection we demonstrate that this is not always the case and introduce a modification to Algorithm~\ref{alg:algorithm-1} that counteracts this issue.

We focus on the following group of objective functions,
\begin{align}
    f_l(x) &= x, \\
    f_q(x) &= x^2, \\
    f_c(x) &= x^3, \\
    f_e(x) &= 2^{10 x / N}.
\end{align}
They map each state to a unique object value, therefore we refer to them as \emph{injective objective functions}.
These injective objective functions are not likely to appear naturally in optimization problems in this simple form.
However, they serve the purpose of demonstrating an important property of Algorithm~\ref{alg:algorithm-1} that may be encountered when using objective functions not explicitly covered in our manuscript.

\begin{figure}[htb]
    \centering
    \includegraphics[width=\columnwidth]{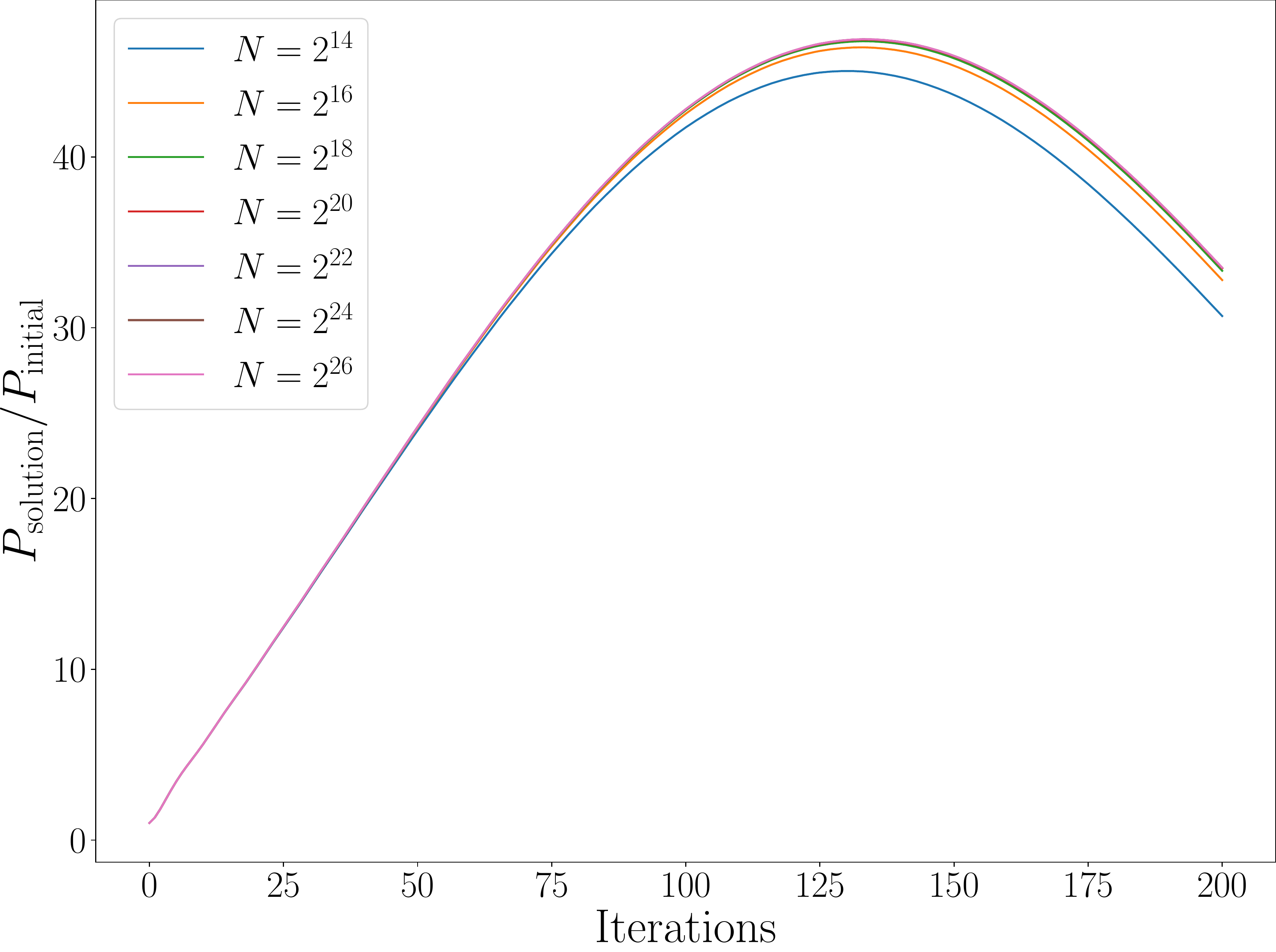}
    \caption{Relative probability to initial uniform probability of measuring the solution state of objective function $f(x) = x^2$ with input size ranging from $N = 2^{14}$ to $N = 2^{26}$. $O_\phi(k)$ for each line is constructed using the same $k$ which is optimal or very close to optimal. Different colors denote different size in state space. The highest probability relative to initial probability approach a certain value when $N$ grows. Noted that the optimal state $\ket{N-1}$ for each line is different but the objective functions are the same.}
    \label{fig:injective-varying-space-size}
\end{figure}

The primary quantity that we are concerned about in this subsection is the relative amplified probability of the optimal solution, $P_{\text{solution}}$, with its initial value, $P_{\text{initial}}=1/N$, before the first iteration of Algorithm~\ref{alg:algorithm-1}.
In particular, we are interested in determining how the ratio $P_{\text{solution}} / P_{\text{initial}}$ behaves for increasing input size space $N$.
Figure~\ref{fig:injective-varying-space-size} displays the scaling of this probability ratio as a function of the number of iterations of Algorithm~\ref{alg:algorithm-1} for increasing $N$.
Interestingly, we can observe that the ratio $P_{\text{solution}} / P_{\text{initial}}$ converges to a single curve for increasing $N$.
This is in stark contrast to behavior observed for normal, skew normal, and exponential distributions where the optimal solution amplitude could be amplified easily independently of the input space size $N$.

One might suspect that we would see a different picture if we fix the objective value ranges but increase the possible data points in between while keeping the density fixed (sampling more from the same density).
Instead of mapping $\ket{x}$ to $x$, we map $\ket{x}$ to $x/m$ where $m$ is some constant and our total states then become $mN$ instead of $N$.
We would still see the same picture, the highest achievable probability relative to initial $1/N$ approach a certain limit.

Figure~\ref{fig:injective-varying-space-size} shows the behavior of $f_q(x)=x^2$ but identical observations were made for the remaining injective objective functions as well.
In the remainder of this section, we present a modification that overcomes this size-induced limit to amplification.

\begin{figure}[htb]
    \centering
    \includegraphics[width=\columnwidth]{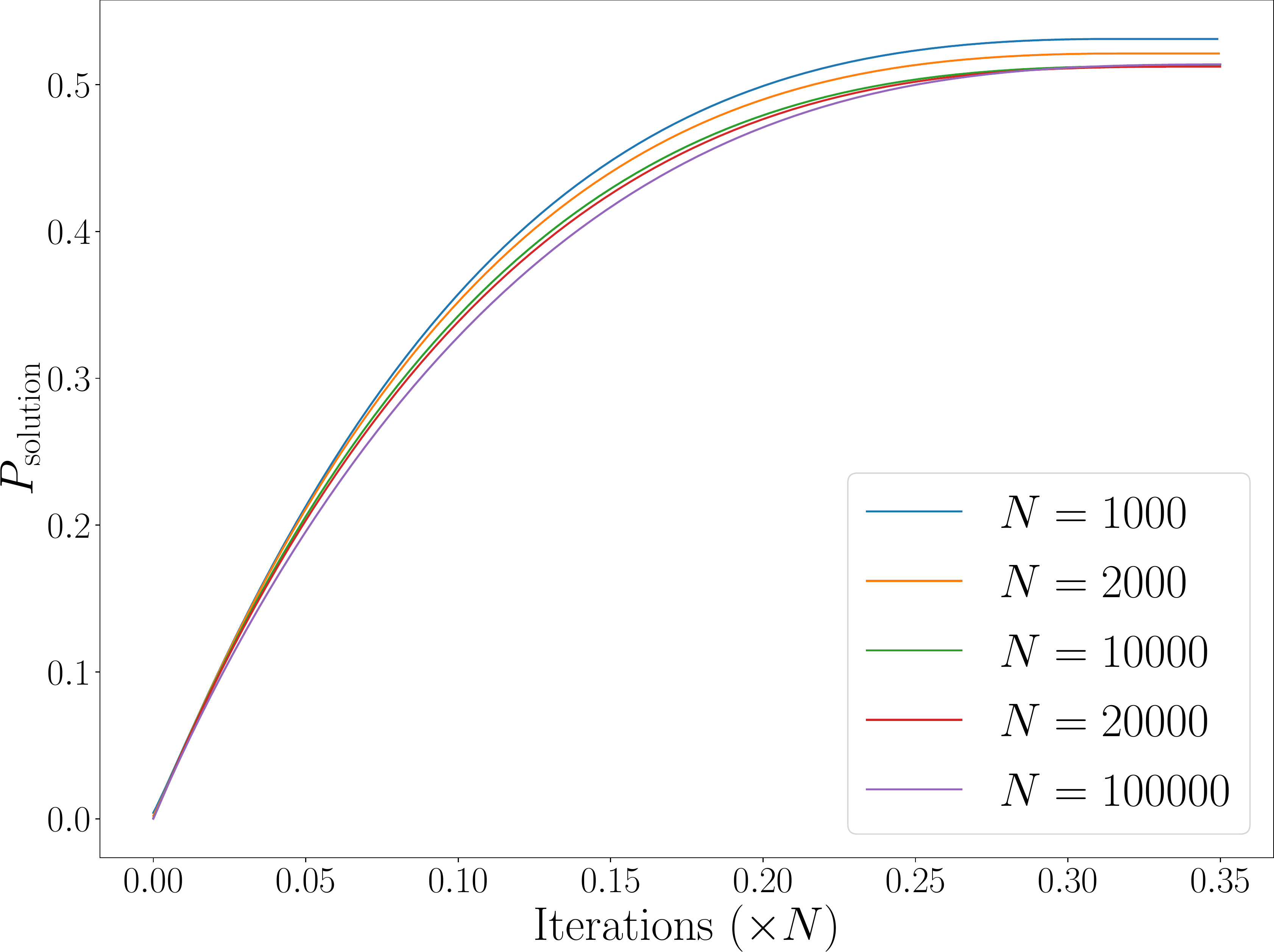}
    \caption{Probability of measuring the solution state of objective function $f(x) = x^2$ with input size ranging from $N = 10^{3}$ to $N = 10^{5}$. The horizontal axis denotes the number of iterations as a fraction of $N$. $O_\phi(k)$ at each iteration is chosen dynamically by greedily picking the $k$ which gives the highest boost in probability.}
    \label{fig:injective-step-varying-k}
\end{figure}

So far we have only considered applying $O_\phi(k)$ with constant $k$ for all iterations of the algorithm. 
There are no real restrictions that this must be the case, and in fact, we can show that by changing the parameter $k$ to the operator $O_\phi(k)$ for each iteration, we can achieve better results and overcome the above mentioned issue. 
Recall that in Fig~\ref{fig:general-trend-sine}, the general trend is that the smaller the phase angle we choose, the shorter the period of the oscillating probability.
This gives us some intuition that a smaller angle can be used in early iterations to quickly boost the amplitude of the solution state then later use a larger angle which allows it to get past the amplification limit of the smaller angle.
An interesting effect of varying the $k$ at each iteration is that it allows the injective objective functions case to get past the negligible boosted probability limit by a single fixed $k$.
We show in Fig~\ref{fig:injective-step-varying-k} for injective families the probability of measuring the solution state at each iteration when using a greedy approach.
$O_\phi$ at each iteration is chosen such that the $P(\ket{N-1})$ is highest by scanning the range of $k$. 

Although the greedy approach to dynamically choose $k$ is not the optimal strategy, it serves as one point in favor of evidence that in general even the injective families of objective functions which cannot be boosted via a fixed choice of $O_\phi(k)$ are boostable with this amplitude amplification technique.

One way to avoid dealing with the complexity when using the subdivided phase oracle $O_\phi$ is to alternate the direction of the rotation for each iteration which can be done by applying $O_\phi(k)$ on the even iteration and $O_\phi(-k)$ on the odd iteration. 
This way the mean vector can be tracked easily as it always stays on the same axis throughout all iterations. 
Although this approach solves the problem of choosing the optimal $k$ since the algorithm performs best when $k$ approaches $0$, the number of iterations requires to amplify the probability grows much quicker than the fixed $k$ method, and the magnitude of the solution state is smaller.
One interesting side effect of this approach is that the states with objective value $f(x)$ higher than the mean get larger while those lower than the mean get smaller. 
This approach of alternating between $O_\phi(k)$ and $O_\phi(-k)$ was explored in detail in~\cite{shyamsundar-nbaa} with the help of an interference pattern created by adding one ancilla qubit to select the rotation direction in superposition which can be thought of as doubling the number of amplitude vectors and splitting them into two groups. Having them rotate in opposite directions forces the mean vector to always stay on the real number line.



%
\section{Quantum Advantage}
\label{sec:advantage}

We have shown that using our modified amplitude amplification algorithm, it is possible to raise the probability of measuring the solution to a significant degree depending on the distribution of objective values.
Since the amplified probability does not reach unity, we still need to resort to classical random sampling after applying the amplitude amplification algorithm.
If we look at Fig~\ref{fig:prob-various-dist}(e), the increase in probability $P_{\text{solution}}$ is faster for the first half than the later half of the iterations required to reach the peak which means that there is an optimal stopping point in which would give us smaller total overall complexity.

In order to properly quantify the speedups against classical scheme, we will use `expected number of trials' to compare quantum and classical schemes.
Let us define: 
\begin{align}
    P_C &= 1/N, \\
    P_Q &= P_{\text{solution}} = \mathrm{P}(\ket{N-1}), \\
    E_C &= N, \\
    E_Q &= t / P_Q,
\end{align}
where $P_C$ ($P_Q$) is the probability of measuring the solution classically (quantumly using Algorithm~\ref{alg:algorithm-1}), and $E_C$ ($E_Q$) is the expected number of queries to get (measure) the solution classically (quantumly), $t$ is the number of iterations or time step count for each trial in the quantum case, and $\mathrm{P}(\ket{N-1})$ is the probability of measuring the optimal solution state $\ket{N-1}$.

We make a reasonable simplification that we treat one iteration of the amplitude amplification algorithm as a single query.
Query complexity is the most used metric to quantify the speedups of Grover's algorithm and amplitude amplification thus we will use this in the same manner.
\begin{figure}[htb]
    \centering
    \includegraphics[width=\columnwidth]{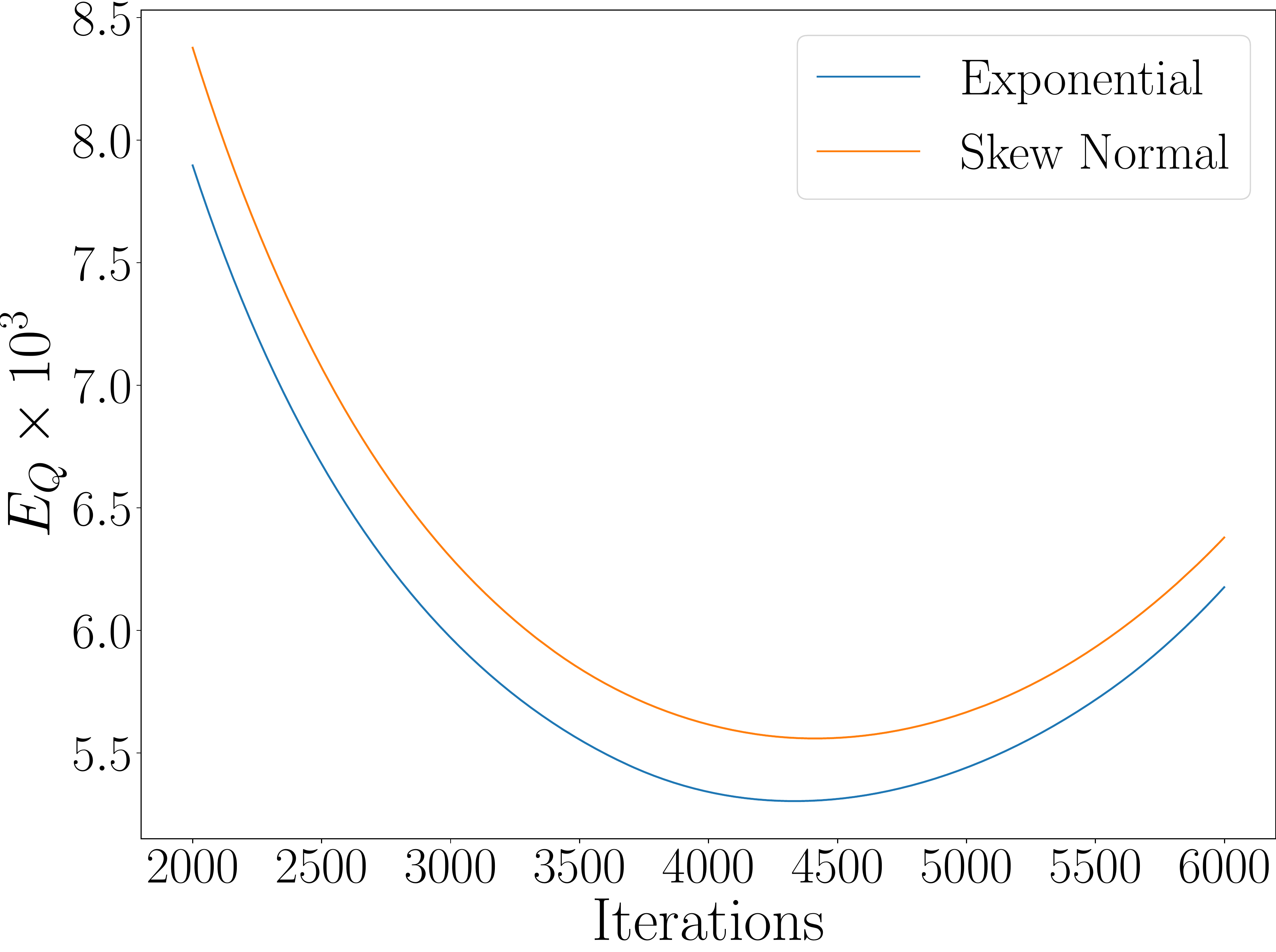}
    \caption{The expected number of queries to measure the solution state via sampling scheme with search space $N \approx 2^{25}$ as a function of the number of iterations. Algorithm~\ref{alg:algorithm-1} is applied for each sampling trial. Note that the minimum of $E_Q$ occurs at a different number of iterations from the maximum of $P_{\text{solution}}$ in Fig.~\ref{fig:prob-various-dist}. Location of the minimum of $E_Q$ indicates the optimal stopping point for the Algorithm~\ref{alg:algorithm-1} after which sampling the state vector $|\Psi_j\rangle$ leads to the best result.}
    \label{fig:tradeoff-advantage}
\end{figure}

As we have seen in the previous sections, the increase in probability does not scale linearly with number of iterations, hence it is not always the optimal strategy to sample after boosting the probability of the solution state to the highest but rather boost it enough such that the expected number of trials is minimum. 
Using the simulations and objective functions mentioned in the previous section, Fig~\ref{fig:tradeoff-advantage} shows the expected number of queries and the trade-off between further amplifying the probability and the number of sampling.

Now that we have defined the expected number of trials to get the optimal solution, we will look at another metric, the success probability of finding the solution.
Since one trial has the success probability $\mathrm{P}(\ket{N-1})$, the success probability when we sample for $p$ trials is
\begin{equation}
    P_{success} = 1 - [1 - \mathrm{P}(\ket{N-1})]^p.
\end{equation}
Using the values obtained in Section~\ref{sec:results}, we see that even a very small number of trials $p$ yields probability of success to be very close to unity.

\section{conclusion}
\label{sec:conclusion}

Even though Amplitude amplification has been known since its discovery decades ago as one of the key quantum tools for achieving an advantage over classical computers, there are still many things to be learned from it. 
In this paper, we showed evidence that the amplitude amplification algorithm in combination with subdivided phase oracle can be used to solve optimization problems, and depending on the structure and distribution of their objective values, an advantage can be gained over classical schemes.
Using a fixed angle oracle $O_\phi(k)$, normal distribution of objective values can be amplified such that the probability of observing the solution state is of a significant degree.
In the case of skew normal and exponential distributions, this probability becomes near unity. 
If changing $k$ at each time-step is allowed, the algorithm is general enough to amplify other distributions of objective values, such as those that are uniform or the injective objective functions. 
The process of choosing an optimal $k$ is non-trivial and needs further study but it has shown promising path in the next step to the more general amplitude amplification of more than two classes of inputs.

There are still so many things to be learned from using the subdivided phase oracle with the diffusion operator, such as the mathematical derivation of the process explained in section~\ref{sec:oracle} or the methods of choosing $k$ when little is known about the problem. 
The visualization technique that we developed and propose in this paper, albeit a simple approach, should be proved useful in designing a better suited diffusion operator to be used with respect to the choice of $O_\phi(k)$ and the distribution.
As with past research on Grover's algorithm, it is known that there is a relationship between the choice of phase shift operator and the angle of diffusion operator for the unstructured search.
It will be interesting to see if the same thing can be derived when the oracle does not separate two groups of states but each of them uniquely. 

From another perspective, the whole process can be viewed as a form of QAOA~\cite{farhi-qaoa, hadfield-qaoa-redefined}, replacing the mixer layer with the diffusion operator~\cite{bartschi-grover-mixer-qaoa}. 
Applying the low-depth QAOA might be a good starting point for selecting $k$ and from there we can vary $k$ using some heuristic algorithm.

Another approach to ease the sensitivity of the choice of $k$ and to improve the potency of the algorithm is to transform the objective function $f(x)$ into another function that exhibits better characteristics.
So far we have only discussed the theoretical aspect of using this algorithm without mentioning the implementation. The embedding of objective function could be referred to the ideas presented in~\cite{gilliam-adaptive-grover, gilliam-binomial-grover} where techniques of directly embedding polynomials can be done without the need for ancilla qubits which might enable the idea of transforming the objective function  without an excessively complex circuit design.



\section*{Acknowledgment}

This work was supported by MEXT Quantum Leap Flagship Program Grant 
Number JPMXS0118067285 and JPMXS0120319794. 



\end{document}